\documentclass[pre,aps,10pt,twocolumn,floatfix,a4paper,nofootinbib]{revtex4-2}
\usepackage[utf8]{inputenc}
\usepackage[T1]{fontenc}
\usepackage[english]{babel}
\usepackage[nopatch=toc]{microtype}
\def\textcite{\cite}
\usepackage{url}
\usepackage{amsmath,array}
\usepackage{graphics}
\usepackage{xcolor}
\definecolor{link}{rgb}{0 0 1}
\usepackage[%
  final,%
  plainpages=false,  
  breaklinks=true,   
  pagebackref=false, 
  hyperindex=false,  
  colorlinks=true,   
  linkcolor=link,    %
  citecolor=link,    %
  menucolor=link,    %
  urlcolor=link,     %
  bookmarks=true,    
  bookmarksnumbered=true,  
  hyperfootnotes=false, 
  pdfcreator={pdfLaTeX},
  pdfstartpage=1]{hyperref}
\usepackage{datetime}
\newdateformat{mydatefmt}{\thisdayofweekname, \ordinaldate{\THEDAY}~\monthname~\THEYEAR}
\def\resub#1{#1}

\def\Vsph{V_\text{sph}}
\def\Vsphatt{V_\text{att}}
\def\Vsphrel{V_\text{rel}}
\def\Vam{V_\text{am}}
\def\Vdeg{V_\text{deg}}
\def\Vtot{V_\text{tot}}

\def\dotR{{\dot R}}

\def\distDV{{d_{VD}}}
\def\maxdistDV{{d_{VD}^\text{max}}}

\def\celsius{\textrm{C}}
\def\micrometer{\mu\textrm{m}}

\def\hour{\textrm{h}}

\def\milligramm{\textrm{mg}}
\def\milliliter{\textrm{mL}}
\def\lenunit{\mathcal{L}}
\def\timeunit{\mathcal{T}}
\def\boxlen{\lenunit}

\graphicspath{{plots}}
\newif\iflowquality
\lowqualityfalse
\begin{document}
\title{Residual semi-crystalline particles released during enzymatic degradation of plastics}
\author{Michael Schindler}
\affiliation{Gulliver, École Supérieure de Physique et Chimie Industrielles, Paris Sciences et Lettres Université, CNRS, Paris 75005, France}
\email{michael.schindler@espci.fr}
\author{Ludwik Leibler}
\affiliation{Gulliver, École Supérieure de Physique et Chimie Industrielles, Paris Sciences et Lettres Université, CNRS, Paris 75005, France}
\mydatefmt
\date{\today~at~\currenttime}

\begin{abstract}
\resub{Enzymatic recycling of plastics is limited by the
presence of semi-crystalline spherulites that are recalcitrant to enzymatic
depolymerization. Depending on the
quality of the waste stream and its treatment history, a large volume fraction of the
material actually remains in form of connected clusters of such spherulites. We
build on a recently published numeric method to predict the number, the connectivity, and the morphology
of these clusters as an outcome of enzymatic degradation. When applied to PET
waste, our method predicts that the resulting aggregates are loosely connected,
``fluffy'' structures with a high surface-to-volume ratio, accompanied by
smaller clusters following a continuous size distribution. By providing a
quantitative framework for understanding the microparticle production during
the depolymerization, these findings should assist choosing a suitable downstream
treatment, such as filtering or flocculation. This work could thus help to
advance the enzymatic depolymerization technologies.}
\end{abstract}
\maketitle
{\sffamily\noindent This document is the unedited author's version of a submitted manuscript
subsequently accepted for publication in 'Macromolecules': Copyright:~2026. The
Authors. Published by American Chemical Society. To access the final published
article, see \url{https://pubs.acs.org/mamobx/}.
}

\section{Introduction}
The enzymatic degradation of polymer waste has emerged as a promising route for
chemical recycling, offering true circularity in the usage of plastic resources
and mild reaction conditions compared to other chemical
methods~\cite{currentPETminireview,ThoHunMey22a,Review26}. Enzymatic
degradation allows \resub{breaking} down polymer chains into short pieces and
finally into monomers, which can then be reused to make fresh virgin plastics,
without the quality \resub{loss} that is observed in other techniques.
Semi-crystalline polymers, however present a challenge to this method: while
amorphous regions are readily broken down by enzymes, \resub{ordered crystalline domains do
not offer sufficient chain mobility and degradable site accessibility, and are
thus depolymerized less
efficiently~\cite{efficientpethydrolases,engineeredPEThydrolases,PatelETAL,Welzel03,Fabre19,compete,pnas17,ThoHunMey22a,ThoAlmMey23,SchThoMey24,pnas21,auclair25}
and about an order of magnitude more slowly. Indeed, crystalline domains
are not perfect monocrystals, often they are spherulites that grew from a nucleation
center. They have lamellar structure, alternating crystalline and
disordered amorphous layers~\cite{SchThoMey24}. Such confined and constrained
amorphous domains within the spherulites are less accessible to enzymes and
more recalcitrant to enzyme-catalyzed depolymerization than amorphous matrix.
Amorphous regions within spherulites are thus depolymerized more slowly than fully
amorphous polymers~\cite{ThoAlmMey23,Fabre19}. In semi-crystalline polymer waste, the
degradation kinetics is therefore coupled to polymer
morphology~\cite{Fabre19,FastCrystPET96,WeiEtal19}. Amorphization pretreatments
alleviate this problem, but amorphization is never complete. Moreover, to be
efficient depolymerization has to be carried out above the glass transition
temperature, at which further crystallization can occur. This concerns in
particular all semi-crystalline polyesters, among which also poly(ethylene
terephthalate)~(PET). Understanding the crystalline morphology and its
evolution during enzymatic treatment is therefore critical for predicting
yields and timescales of the enzymatic degradation process.}

Recent experimental and modelling work~\cite{ourAvrami} has revealed that under
typical conditions the enzymatic \resub{depolymerization} of amorphous matrix competes with
the growth of embedded spherulitic, partly crystalline, domains. \resub{The
presence of both, the amorphous matrix and the spherulites, results from the
pre-treatment of the plastic waste, in particular melting, \resub{rapid cooling,} and mechanical
grinding.} \resub{This competition between depolymerization and spherulite growth then} determines the final yield and the timescale of the
degradation process. The undegraded, residual material represents a significant
fraction of the initial polymer mass. Depending on the quality of the waste
stream and its treatment history, at least $40\%$~of the material actually
remains in form of spherulitic clusters~\cite[figure~12]{ourAvrami}.

Questions naturally turn to how to further process the leftover material.
Beyond its total volume, also the morphology of the residual material plays an
important role: Rather than yielding a uniform reduction in particle size, the
process generates a heterogeneous population of clusters of crystalline
spherulites that are eventually released into the surrounding solution.
Moreover, a closer look on these spherulites reveals that they are not totally
crystalline but consist of lamellae or similarly ordered stacks, with amorphous
material between them~\cite{DiLorenzo24,Wunderlich03,ourAvrami}. \resub{For
example in reference~\cite[Figure~10]{ourAvrami}, efficient amorphization
pretreatment yielded an initial crystallinity degree as low as~10\%, which
corresponds to a spherulite volume of already~35\%.} The small-scale amorphous phase
\resub{within the spherulites}
can be depolymerized, but it is less accessible to enzymes than the bulk
amorphous matrix and has less mobility, and it is thus depolymerized \resub{on
a larger timescale~\cite{Welzel03}. It is thus useful to
differentiate between a ``first stage'', in which the bulk amorphous matrix is
depolymerized, and in which partly crystalline spherulites grow; and a ``second
stage'', in which also the amorphous material within the spherulites is
depolymerized~\cite[figure~12]{ourAvrami}. The two stages are not strictly
successive in time, they refer to the different timescales involved.}

The rate at which \resub{the slow ``second-stage''} depolymerization happens,
strongly depends on the morphology of the clusters \resub{that result from the
faster ``first stage'': If the clusters expose much surface to the enzymes, for
example} if they are rather fractal objects with thin local structures, they
are depolymerized at a higher rate than if they were a dense compact object,
exposing little surface. \resub{Knowing the morphology of this material is thus
interesting and important. In the present paper we analyze the morphology of
the clusters resulting from the ``first stage''.}

The morphology of the undegraded polymers \resub{might influence} the overall
design of the recycling process. There are various decisions to take. Either
the crystalline spherulites are kept in the enzyme solution until they are also
degraded on a long or very long timescale \resub{(the \emph{waiting strategy})}; or they have to be flocculated or
filtered out and treated \resub{similarly} as the initial plastic waste, that
is melted, cooled, \resub{milled, and depolymerized} \resub{(the \emph{filtering strategy})}. In order to take such a
design decision, it is important to know the size distribution of the
\resub{spherulites produced} (for the filtering strategy) and how much
surface they potentially expose to enzymes (for the waiting strategy).
\resub{For example, many very small spherulites in the outcome propose themselves for
the waiting strategy, whereas compact big particles require to be filtered and
subjected to another round of the whole process.}
\begin{figure}%
  \centering
  \includegraphics{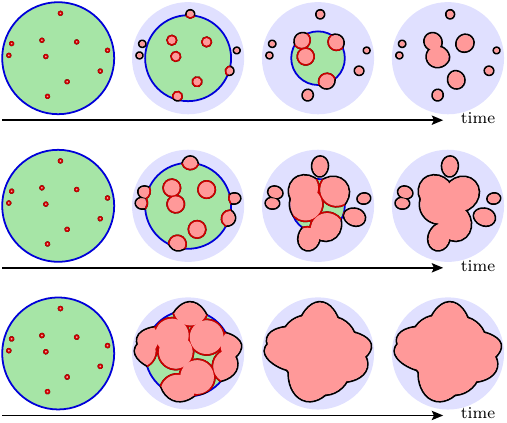}%
  \caption{Possible outcomes of the enzymatic degradation of a plastic
  particle, for different processing conditions: many small spherulites (top
  row), or various sizes of spherulite clusters (middle row), or a single
  compact cluster of spherulites (bottom row). The amorphous matrix (green) is
  degraded, the spherulites (red) grow (and possibly nucleate) within the
  amorphous matrix, finally remaining as connected clusters of egg-shaped
  objects (black curves).}\label{fig:volsurf_evolv}
\end{figure}%

Despite their importance for process optimization, the geometric and
statistical properties of these residual clusters have received limited
attention~\cite{PanHan09,Hedenqvist22,Fabre19}. Key questions remain: How does
the density and growth rate of spherulites determine cluster morphology? Under
what conditions do spherulites remain isolated versus forming large aggregates?
What surface area do these clusters expose for further enzymatic attack?
Answering these questions requires connecting the microscopic parameters of
spherulite nucleation and growth to the macroscopic observables of cluster size
distributions and surface properties.

Here we address these questions using a geometric model that treats degradation
as the coupled evolution of a shrinking amorphous sphere containing randomly
distributed, growing spherulites. Building on our previous
work~\cite{ourAvrami}, which focused on the total amount of degraded and
remaining volumes, we now characterize the residual clusters themselves. By
solving the model equations numerically and tracking the connectivity of
overlapping spherulites, we identify when clusters detach from the degrading
particle and determine their volume, surface area, and internal structure.
Applying \resub{these} numerics to the parameters we obtained \resub{in
reference}~\textcite{ourAvrami} for the experimental degradation of PET from
different waste sources, allows us to make a prediction for the total amount
and for the distribution of structure and size of the leftover crystalline
clusters. \resub{Since experiments suggest that the depolymerization of
spherulites is much slower than that of the amorphous
matrix~\cite[figure~12]{ourAvrami}, we neglect it in the model. This work thus
answers the question how far depolymerization can go if only the amorphous
matrix is degraded. It does not explicitly describe the depolymerization of
inter-lamellar amorphous material in clusters. Instead, it provides the
morphology of these clusters, which is useful for estimating their further
degradation.}

Our results demonstrate how the operating parameters of enzymatic degradation
-- temperature (which controls spherulite growth rate), residence time, and
initial particle size -- map onto a percolation phase diagram that predicts
residual cluster morphology. This connection between process conditions and
residual waste structure provides a much simpler but still quantitative
framework for optimizing multi-stage degradation protocols: conditions which
maintain subcritical spherulite density yield residues in small clusters only,
amenable to rather quick secondary treatment, whereas supercritical conditions
produce compact aggregates requiring repeated processing strategies such as
mechanical separation and re-melting.

Our results further demonstrate how to predict cases in which remains a big
structured cluster of spherulites. We develop a reduced model to predict the
time of its formation, which marks also the end of the degradation process.

More broadly viewed, this work illustrates how concepts from soft matter
physics -- percolation transitions, surface-to-volume scaling, extreme value
statistics, and the statistical mechanics of random geometric structures --
illuminate practical problems in sustainable materials processing. The
geometric model we employ is sufficiently simple to yield analytical insights
while capturing the essential physics of competing degradation and
crystallization. Extensions to more complex scenarios, including nucleation,
polydisperse spherulite populations, non-uniform distributions, and anisotropic
growth, represent natural next steps in developing predictive models for
polymer degradation.

The paper is organized as follows: Section~\ref{sec:model} reviews the
geometric model and establishes notation. Section~\ref{sec:clusters} presents
numerical solutions of the model, including the time-resolved release of
spherulite clusters into the solute. It provides the distributions of cluster
sizes and their surface properties. In sections~\ref{sec:percolation}
and~\ref{sec:extreme} we develop two sub-models which predict the two relevant
timescales of the process, one for the formation of a dominant cluster, the
other for its ``release'' into the solute. Section~\ref{sec:exp} applies the
theory to parameters obtained from experimental data.

\section{Summary of the geometrical model}
\label{sec:model}
%
\begin{figure}%
  \centering
  \includegraphics{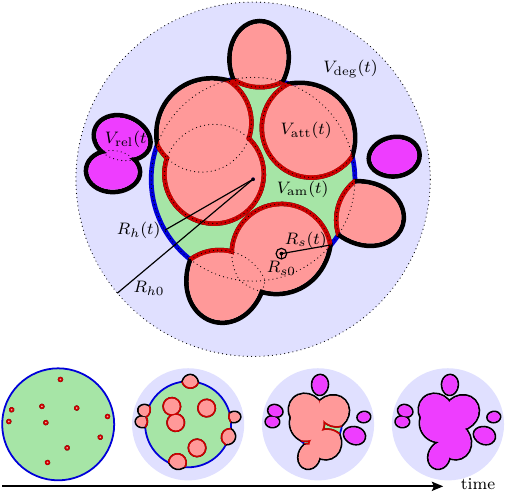}%
  \caption{Top: the four different geometric volumes used in the model, at a
  given time~$t$: Already degraded volume~$\Vdeg(t)$ (light blue), the
  amorphous matrix~$\Vam(t)$ (light green), the volume of spherulite clusters
  still attached to the matrix~$\Vsphatt(t)$ (light red), and the volume of
  spherulite clusters not attached anymore~$\Vsphrel(t)$ (light purple). The
  thick curves show the interfaces between these volumes, and the dotted curves
  indicate spherical shapes. Bottom: the same volumes at consecutive
  times.}\label{fig:volsurf}
\end{figure}%
In reference~\textcite{ourAvrami} we presented a geometrical model for the
enzymatic degradation of plastic waste, where the waste is thermally
pre-treated and shred into small particles. The main issue addressed by the
model is that these polymer particles are semi-crystalline: They contain partly
crystalline spherulites which are degraded much more slowly. The nucleation and
growth of spherulites inside the otherwise amorphous plastic particles is a
result of the chosen degradation conditions. In particular, the reaction
temperature must be high enough such that the polymer chains in the amorphous
parts are sufficiently mobile for the enzymes to cleave them. At this
temperature crystalline spherulites nucleate and grow in time. The overall
degradation is then a competition between \resub{depolymerization} of the amorphous
matrix at its surface and the growth of embedded spherulites.

We take spherical geometry for both the degrading particle and for the growing
spherulites. The degradation of the amorphous matrix is modelled by the
shrinking of a sphere, having at time~$t$ the radius
\begin{equation}
  \label{eq:Rht}
  R_h(t) = R_{h0} + \dotR_h t, \quad\text{with constant $\dotR_h < 0$.}
\end{equation}
The radius decreasing linearly in time reflects that the volume change of
amorphous matrix material is proportional to the area of surface it exposes to
enzymes.

Each spherulite has a given initial radius and grows at constant growth rate
where it is in contact with the amorphous matrix. For the sake of simplicity we
take the same initial radius and the same growth rate for all spherulites, and
we take the spherulites' radius of curvature to grow linearly in time,
\begin{equation}
  \label{eq:Rst}
  R_s(t) = R_{s0} + \dotR_s t, \quad\text{with constant $\dotR_s > 0$.}
\end{equation}
Initially, the spherulite positions are uniformly randomly distributed within
the amorphous sphere, such that they do not stick out. Their number is denoted
by~$N_s$ and their density of centers (number per volume) by~$N_s'$.

As the radii evolve in time, the embedded spherulites can come into contact
with each other, and the amorphous matrix is degraded from the outside, such
that spherulites can start to stick out. Spherulites are modelled to be
non-degradable by enzymes, such that the part of the spherulites outside the
amorphous matrix does not change anymore. The combined dynamics finally leaves
clusters of spherulites as a remainder. The shapes at some intermediate times
are shown in figures~\ref{fig:volsurf_evolv} and~\ref{fig:volsurf}. The
characteristic egg-like shape of a spherulite, with its long axis oriented
towards the center of the particle, is the result of intersecting a growing
with a shrinking sphere. Figure~\ref{fig:volsurf} depicts volumes of three
different phases: The amorphous matrix~$\Vam(t)$ in light~green; the
volume~$\Vdeg(t)$ of matrix having been degraded up to time~$t$ in light~blue;
and the spherulite volume $\Vsph(t)$ which has some (partly) crystalline
structure. These three volumes always add up to the initial volume $\Vtot :=
4\pi R_{h0}^3/3$, which will serve to \resub{normalize} them,
\begin{equation}
  \label{eq:Vtot}
  \Vam(t) + \Vsph(t) + \Vdeg(t) = \Vtot.
\end{equation}
Within the volume~$\Vsph(t)$ of the spherulite/crystalline phase, we
differentiate between those spherulite clusters having been released into the
surrounding and those clusters still \resub{attached to the} amorphous matrix --
either embedded in it, or partly sticking out. The former have
volume~$\Vsphrel(t)$ and are drawn in light~purple in figure~\ref{fig:volsurf},
whereas the latter have volume~$\Vsphatt(t)$ and are drawn in light~red. They
satisfy
\begin{equation}
  \label{eq:Vsph}
  \Vsph(t) = \Vsphatt(t) + \Vsphrel(t).
\end{equation}
Figure~\ref{fig:volsurf} further shows the interfaces between the different
volumes. The rate at which amorphous volume is converted into spherulite volume
is proportional to the area of the amorphous--spherulite interface (dark~red in
figure~\ref{fig:volsurf}). This quantity is -- up to a constant scaling factor
corresponding to the internal degree of crystallinity of the spherulites --
accessible by experimental
techniques~\cite{ourAvrami,FastCrystPET96,CrystPET17,ourPNAS}.

At the same time as spherulites are shaped by the receding amorphous matrix,
they block access of enzymes to this matrix and thus effectively reduce the
degradation rate. The degradation of the amorphous volume is thus proportional
to the area of the exposed amorphous--solute interface (thick dark~blue curves
in figure~\ref{fig:volsurf}).
\begin{figure}%
  \centering
  \includegraphics{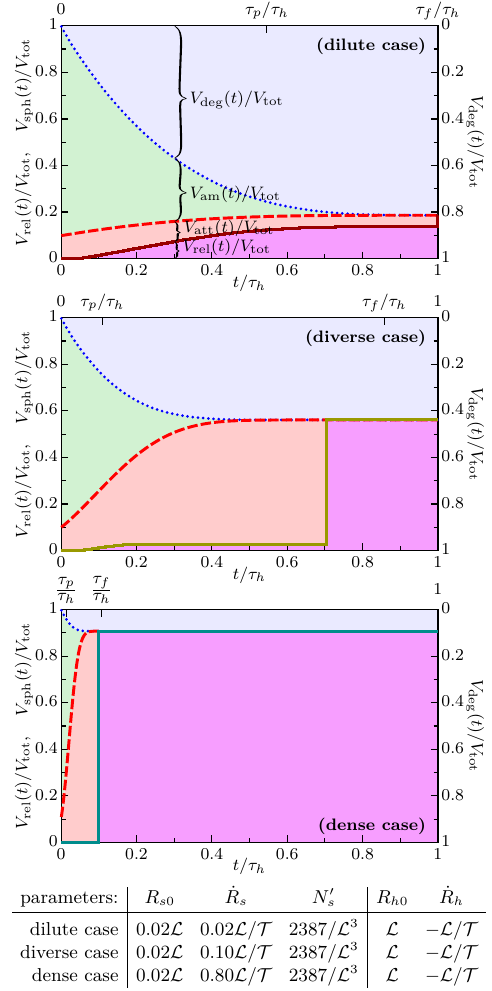}%
  \caption{Evolution in time of the four volumes $\Vsphrel(t)$, $\Vsphatt(t)$,
  $\Vam(t)$, $\Vdeg(t)$. As guides to the eye, we use the same color code for
  these volumes as in figure~\ref{fig:volsurf}. The dashed red curves and the
  solid stepwise lines refer to the left axes, the blue dotted curves refer to
  the right axes. The characteristic percolation time ($\tau_p$) and the
  characteristic final time of filling ($\tau_f$) -- see below, in
  equations~\eqref{eq:taup} and~\eqref{eq:taur} -- are marked on the upper time
  axis of each graph. Top graph:~spherulites remain individual or form small
  clusters (dilute case). Middle graph:~diverse sizes of smaller clusters are
  released, with one bigger clusters at late time. Bottom graph:~the growing
  spherulites remain attached to the amorphous phase only for short time and
  are released as one big cluster (dense case). The table gives the model
  parameters used for the graphs.}\label{fig:release}
\end{figure}%
The characteristic timescale of degradation is given by
\begin{equation}
  \label{eq:tauh}
  \tau_h := R_{h0} \bigm/ |\dotR_h|,
\end{equation}
which is the time a purely amorphous particle takes to degrade. We denote units
of length and time by $\lenunit$ and $\timeunit$, respectively.

\subsection*{Finally degraded volume}

Solutions to equations~\eqref{eq:Rht}--\eqref{eq:Vtot} have been presented in
reference~\textcite{ourAvrami} in form of the time evolution of the volumes
involved. Figure~\ref{fig:release} shows such solutions for three prototypical
choices of parameters. The main quantities of interest in
reference~\textcite{ourAvrami} were the final yield, i.e.~the amount of degraded
material, and the timescale on which it is obtained. We discussed the influence
of the parameters $R_{h0}$, $R_{s0}$, $\dotR_h$, $\dotR_s$, and $N_s'$.

\subsection*{Remaining spherulite clusters}

Since we model the spherulites to be inert to enzymes, and the spherulites grow
only inside the amorphous sphere, where they are in contact with the amorphous
matrix. Thus, spherulite clusters, consisting of partly overlapping
spherulites, grow while at least one of their member spherulites is still
attached to the matrix. Once the last member spherulite has detached from the
matrix, the cluster of spherulites is considered \resub{as ``released'' and diffusing}
away from the particle.

\section{Released clusters of spherulites}
\label{sec:clusters}
The numerical solution of the model equations, applied to explicit sets of
spherulite positions, \resub{gives access to} much more detailed information than we
presented in reference~\textcite{ourAvrami}. There we provided the amount of
degraded volume and the total amount of spherulite volume. The numerical method
uses adapted Delaunay/Voronoi tessellations to integrate volume and surface of
\emph{all individual spherulites}, and also to determine which of them overlap.
We thus have access to their connectivity and can identify all connected
clusters of spherulites~\cite{cgal:union_find}. For each such cluster we
determine whether it is still attached to the amorphous matrix, that is whether
one of its spherulites overlaps with the shrinking sphere. Once a cluster has
detached, it is considered ``released'', and we have access to its volume and
to its surface, by adding up the contributions from all spherulites taking part
in the connected cluster.

\resub{The numerical technique thus allows} to model the creation of spherulite clusters during
the degradation process. Figure~\ref{fig:release} shows the evolution in time
of the four different volumes in the model: The blue dotted curves, referring
to the right axes, show the volume~$\Vdeg(t)$ having been degraded up to
time~$t$. The red dashed curves, referring to the left axes, show the volume of
spherulites, $\Vsph(t)$. Thus, the volume of amorphous matrix, $\Vam(t)$ is the
vertical distance between blue and red curve -- see the vertical braces in the
top panel of the figure. The spherulite volume $\Vsph(t)$ is split into two
parts, namely $\Vsphrel(t)$, the volume of spherulite clusters having detached
from the amorphous matrix and been released into the solute, and $\Vsphatt(t)$
the volume of spherulites still connected to the amorphous matrix at time~$t$,
either directly or indirectly as part of a cluster still being attached. The
identification of the four volumes in figure~\ref{fig:release} has been eased
by using a background coloring scheme using the same colors as in
figure~\ref{fig:release}.

In figure~\ref{fig:release} the two types of spherulite volumes, the released
clusters and the attached ones, are separated by lines consisting of many
discrete steps. At each step a spherulite cluster is released from the
particle. There are many small steps, hardly visible individually, and a final
(big) step which is the ``release'' of the remaining spherulite cluster, when
all amorphous matrix has been degraded.

In the top panel of figure~\ref{fig:release} the overall degradation is very
good, and only $19\%$~of the initial material is finally found in form of
spherulites. The spherulites are continuously released as very small clusters
until the end of the process. This case corresponds to small and slowly growing
spherulites. In the middle panel spherulites grow faster, leading to the
release of a few very small clusters in the beginning; the main volume,
however, is connected to a main cluster until the end of the depolymerization
at quite late time. The final cluster has around~$50\%$ of the initial volume.
In the bottom panel the spherulites grow so fast that they clump together
quickly and form one dense cluster after short time. There are no small
clusters, and only a few percent of the matrix material could be degraded.
\begin{figure}%
  \centering
  \includegraphics{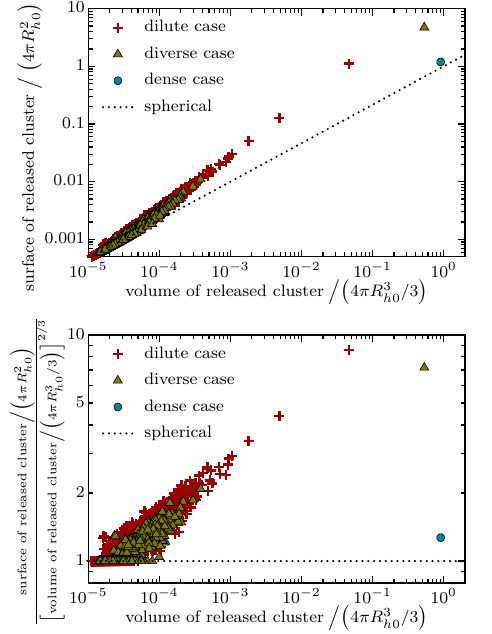}
  \caption{Top graph: Volume and surface of every cluster formed in the
  evolutions shown in figure~\ref{fig:release}. Bottom graph: The same data,
  plotted on a different vertical axis which shows how ``sphere-like'' the
  cluster is. Numeric values are given in table~\ref{tab:cluster_measures}.}%
  \label{fig:cluster_measures}
\end{figure}%
\begin{table}[b]%
  \centering\noindent
  \iflowquality
\else
    \def\hsep{\rule{5pt}{0pt}}
    \def\vsep{\rule{0pt}{10pt}}
    \begin{tabular}{r<{\hsep}|c|c|c}
       & dilute & diverse & dense \\\hline
      $N_s$ & 9939 & 9939 & 9939 \vsep\\
     number of clusters & 3700 & 738 & 1 \vsep\\
     \rule{0pt}{16pt}
     $\displaystyle\sum\limits_c\frac{\text{volume of cluster}~c}{4\pi R_{h0}^3/3}$ & 0.19 & 0.56 & 0.91 \\
     \rule{0pt}{16pt}
     $\displaystyle\sum\limits_c\frac{\text{surface of cluster}~c}{4\pi R_{h0}^2}$ & 5.9 & 5.6 & 1.2 \\
     \rule{0pt}{20pt}
     $\displaystyle\frac{\sum_c\frac{\text{surface of cluster}~c}{4\pi R_{h0}^2}}%
                        {\bigl[\sum_c\frac{\text{volume of cluster}~c}{4\pi R_{h0}^3/3}\bigr]^{2/3}}$ & 18.2 & 8.2 & 1.27 \\
     \rule{0pt}{20pt}
     $\max\limits_c\frac{\frac{\text{surface of cluster}~c}{4\pi R_{h0}^2}}%
                        {\bigl[\frac{\text{volume of cluster}~c}{4\pi R_{h0}^3/3}\bigr]^{2/3}}$ & 8.6 & 7.2 & 1.27 \\

     \rule{0pt}{20pt}
     $\displaystyle\max\limits_c\frac{\text{volume of cluster}~c}{4\pi R_{h0}^3/3}$ & 0.047 & 0.53 & 0.91 \\
     \rule{0pt}{20pt}
     \parbox[b]{3.8cm}{maximal number of\hfill\hbox{}\\[-1.2\baselineskip]\hbox{}\hfill spherulites in a cluster} & 1533 & 8648 & 9939
    \end{tabular}
  \fi
  \caption{Quantitative measures of the spherulite clusters shown in
  figure~\ref{fig:cluster_measures}.}%
  \label{tab:cluster_measures}
\end{table}

Figure~\ref{fig:cluster_measures} shows volume and surface of every released
cluster, including the final cluster, for the three cases given in
figure~\ref{fig:release}. The ``dilute'' case is dominated by individual
sphere-like spherulites; there are also many small clusters of spherulites
partly overlapping. In the opposite, ``dense'' case only one cluster remains.
Its shape is close to a sphere, having only little extra surface, which comes
from spherulites bulging out a bit. In the intermediate case of ``diverse''
cluster sizes, we find again small ones, but there is also a dominating
cluster, which is loosely connected and has a large surface. It contains about
half of the initial volume. Table~\ref{tab:cluster_measures} provides
quantitative information about the clusters, in particular their number, the
sum of their volumes and of their surface areas. \resub{Also, the} largest clusters are
quantified there.

In order to obtain a clearer idea of how much more surface these clusters
expose, in figure~\ref{fig:cluster_measures} we plot the same data in two
different ways. In the lower panel, the vertical axis provides a
surface-to-volume ratio for each cluster, which compares the cluster's surface
with the surface of a sphere having the same volume as the cluster. This
measure gives an idea how much excess surface the cluster has due to its
complex structure. We find that in the ``dense'' case volume and surface are
close to those of a sphere, its surface is only $1.27$~times larger, see again
table~\ref{tab:cluster_measures}. It is thus a compact, round object; the
lowest row in figure~\ref{fig:volsurf_evolv} gives a good idea of how the
``dense'' final cluster might look like.

Very differently, in the ``diverse'' case we find much more surface, with a
total surface-to-volume ratio of~$8.2$. Already the dominating cluster has a
ratio of~$7.2$.

Finally, in the ``dilute'' case we find only small clusters, the largest
cluster volume has only $4.7\%$~of the total cluster volume. The many small
clusters together have huge excess surface, with a surface-to-volume ratio
of~$18.2$. This is not surprising, since we have $3700$~clusters, formed from
the initial $9939$~spherulites, thus on average only three spherulites per
cluster. Nevertheless, there are also bigger clusters, up to $1533$~spherulites
for the largest one. This cluster has a similar structure as the dominating one
in the ``diverse'' case, its surface-to-volume ratio is~$8.6$.

The dramatic increase of surface, relative to volume, of small or loosely
connected clusters is a true advantage for \resub{a possible} second stage of degradation.
With eight or eighteen times more surface even a five times slower degradation
can be useful.
\begin{figure}%
  \centering
  \includegraphics{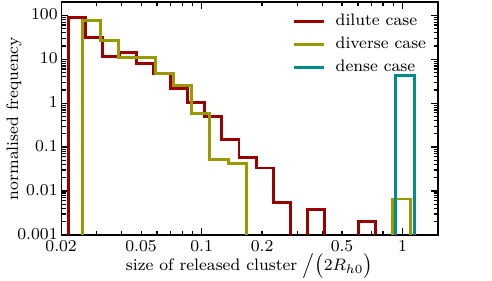}
  \caption{\resub{Histogram of sizes of the clusters formed in the evolutions shown in
  figure~\ref{fig:release}. Here, ``size'' refers to the maximal extent along
  the three Cartesian coordinate axes.}}%
  \label{fig:cluster_sizes}%
\end{figure}%

\resub{Figure~\ref{fig:cluster_measures} gives an indication how compact a
cluster is, based on its volume and surface. It gives a rough idea of the
cluster's size -- but rather indirectly. In figure~\ref{fig:cluster_sizes} we
show a histogram of the sizes of these clusters. For complex objects made of
several spherulites, the term ``size'' is not as well-defined as are volume and
surface. Here, we measured the extent of the cluster parallel to each Cartesian
coordinate axis, that is the smallest and the largest Cartesian coordinate of
all its spherulites. The ``size'' referred to in figure~\ref{fig:cluster_sizes}
is the largest such extent.}

\resub{In figure~\ref{fig:cluster_sizes} we see the same qualitative picture
as in figure~\ref{fig:cluster_measures}: The ``dilute'' case results in a
continuum of clusters, where larger clusters are increasingly rare. The
``various'' case exhibits a discontinuous distribution of sizes, with small
clusters and a large one, essentially of the same size as the initial particle.
The gap between the sizes is well visible. The ``dense'' case results only in
large clusters, again of the same size as the initial particle.}

\section{Focus model: Predicting the percolation time}
\label{sec:percolation}
\resub{In figures~\ref{fig:release}, \ref{fig:cluster_measures} and~\ref{fig:cluster_sizes} we find a crossover
between two extremes}, one being mainly isolated spherulites, or very small
clusters of them, the other being a single big cluster of nearly all
spherulites. We can \resub{rationalize} this result using a simplification that is
known in the scientific literature of percolation thresholds as the ``continuum
percolation model''\footnote{Various names have been given to the model, such
as ``off-lattice percolation'', ''fully penetrable spheres'', ''overlapping
spheres'', the ``Swiss-cheese model'', or the ''Poisson blob model.''}
\cite{QuintanillaTorquato96,Kratky78,StaufferAharony94,MeesterRoy96}. There,
many spheres are randomly placed in infinite or periodic space. Mutual overlap
of spheres defines a ``connectivity'' between them, such that ``clusters'' of
connected spheres can be identified. The main question addressed in the context
of percolation is whether there is a dominant cluster.

\begin{figure}%
  \centering
  \includegraphics{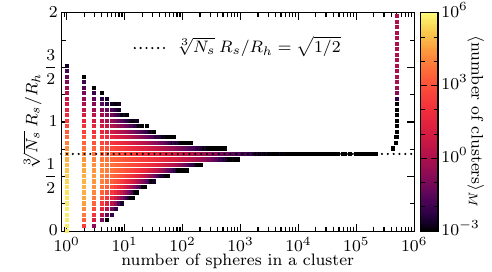}%
  \caption{The percolation problem: Numerically determined frequencies of
  cluster sizes, when many ($N_s{=}5{\times}10^5$, fixed) small spheres
  (radius~$R_s$, varying) are placed uniformly randomly inside one big sphere
  (radius~$R_h{=}1$, fixed). The numbers of clusters are averaged over many
  ($M{=}10^3$, fixed) repetitions of the random placing. Dots are drawn where
  at least one cluster comprising the given number of spheres has been
  observed. Their color indicates how many such clusters have been observed,
  averaged over the repetitions. The dotted line is an estimate for the
  percolation threshold.}\label{fig:percolation}
\end{figure}

Let us simplify the setting of our model and neglect the time evolution of
spherulites and of particles. We do numerical calculations of the percolation
test, adapted to our setting of spherulites within a particle: In three spatial
dimensions we randomly place $N_s$~small spheres of radius~$R_s$ in a big
sphere of radius~$R_h>R_s$. Their spatial distribution is chosen such that the
centers are uniformly distributed inside the large sphere. We then determine
the connectivity of the small spheres and count for each connected cluster the
number of involved small spheres. We further count how many clusters we have
for a given number of involved spheres. The whole process is repeated~$M$ times
to obtain better statistics, and it is repeated for various values of~$R_s$.
Figure~\ref{fig:percolation} presents the result: For every observed number of
spheres in a cluster the figure shows a little square. Its color encodes how
often this cluster size has been observed, averaged over the repetitions.
\begin{figure}%
  \centering
  \includegraphics{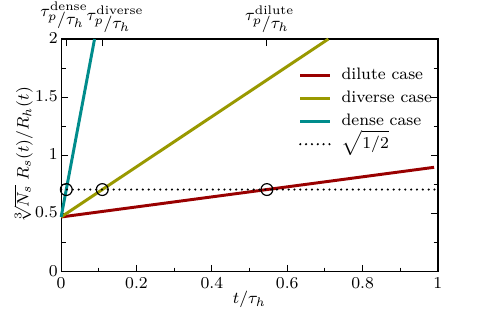}%
  \caption{The percolation indicator~$\sqrt[3]{N_s}R_s(t)/R_h(t)$ as a function
  of time, for the same parameters as for the curves in
  figure~\ref{fig:release}. The times when they cross the percolation threshold
  (dotted line) are marked by circles and on the upper time
  axis.}\label{fig:param_paths}
\end{figure}%

The percolation \resub{transition} is very well visible in figure~\ref{fig:percolation}:
There is a threshold value near $\sqrt[3]{N_s}R_s/R_h \approx \sqrt{1/2}$, see
also reference~\textcite{Torquato12}. Below the threshold, for small or few
spherulites, cluster sizes behave monotonously, small clusters are very
numerous, bigger clusters less numerous, and there \resub{is no dominating cluster}.
Above the threshold, \resub{one cluster dominates, combining nearly all
spheres, accompanied by only a few additional small clusters}.

The percolation model allows to understand the difference between the three
cases in \resub{figures~\ref{fig:release}, \ref{fig:cluster_measures} and~\ref{fig:cluster_sizes}} qualitatively
and quantitatively. In figure~\ref{fig:param_paths} we draw the parameters as a
function of time, such that we can trace their combination relevant for the
clustering, namely $\sqrt[3]{N_s}R_s/R_h$ over time. These parameter functions
are linear, essentially showing the growth of the spherulite radius~$R_s(t)$.
The times when the parameter lines cross the percolation threshold are marked
with circles in figure~\ref{fig:param_paths}. They are given by resolving the
percolation condition
\begin{equation}
  \label{eq:percolation_condition}
  \sqrt[3]{N_s}R_s/R_h=\sqrt{1/2}
\end{equation}
for the time, yielding the characteristic time of percolation
\begin{equation}
  \label{eq:taup}
  \tau_p := \frac{1}{\dotR_s}\biggl(\frac{\sqrt{1/2}}{\sqrt[3]{N_s'4\pi/3}}-R_{s0}\biggr).
\end{equation}
Inserting the parameters of figure~\ref{fig:release}, we obtain
\begin{equation}
  \iflowquality
\else
    \def\hsep{\rule{5pt}{0pt}}
    \def\vsep{\rule{0pt}{10pt}}
    \begin{array}{r<{\hsep}|>{\hsep}c}
                          & \tau_p/\tau_h \\\hline
      \text{dilute case}  & 0.55\phantom{0} \vsep\\
      \text{diverse case} & 0.11\phantom{0} \\
      \text{dense case}   & 0.014 \rlap{.}
    \end{array}
  \fi
\end{equation}
These values are also indicated on the upper time axes of the graphs in
figure~\ref{fig:release}. In the top two graphs we see that $\tau_p$~indicates
the typical timescale when small spheres add up to a considerable volume. This
is precisely the setting of the percolation approximation.

With the percolation model in mind, it becomes an evident fact that for an
initial volume of spherulites higher than~$0.35\%$, more precisely for
$\Vsph(0)/\Vtot > 2^{-3/2}$, the dilute case is impossible. There will always
be a dominant cluster, either densely packed or fluffy.

\section{Focus model: Predicting the time when depolymerization ends}
\label{sec:extreme}
In the previous section we presented a reduced effective model for the
percolation time~$\tau_p$. Here we proceed to develop another reduced effective
model focusing on the ``release'' time of a dominant remaining cluster,
$\tau_f$. This time marks the end of the depolymerization of the particle and
is well visible as a pronounced final step in each graph of
figure~\ref{fig:release}. In the detailed numerics of figure~\ref{fig:release},
the time of ``release'' is defined to be the moment when the shrinking sphere,
which represents the amorphous matrix, exposes no surface to the
enzymes anymore. \resub{Some enclosed cavities of amorphous
matrix may still be left, but they are small and will disappear} shortly after. Effectively,
the time of release nearly coincides with the time when the sphere representing
the amorphous matrix is completely filled by spherulites. We are thus looking
for the formation time of a dense core of the remaining cluster.

Our reduced model should therefore answer the question at what (growing)
radius~$R_s(\tau_f)$ spherulites, which have been placed randomly in space at a
given density (of centers)~$N_s'$, fill space completely.

We consider again a Delaunay tessellation, in which every vertex carries one
spherulite center. A Delaunay cell is then a tetrahedron (simplex) having four
such Delaunay vertices at its corners, and it corresponds to one Voronoi vertex
of the dual tessellation. The dual (Voronoi) vertex has the same
distance~$\distDV$ to all four vertices of the Delaunay cell. This means that
the growing spherulites, centered in the corners of the Delaunay tetrahedron,
will meet in that very dual Voronoi point at the moment they completely fill
the Delaunay tetrahedron. At that time all four spherulite radii will be equal
to the distance $\distDV$~of this cell. Since the positions of spherulites are
random variables, also the distance~$\distDV$ and the complete-filling time are
random variables, for each Delaunay cell.

The latest closure of any point in the whole tessellation will take place at
the Voronoi vertex which has the largest distance to its Delaunay neighbors.
This is the respective maximum of all $\distDV$ in this tessellation,
$\maxdistDV$, which is also a random variable. We will ask for its distribution
and its \resub{mean value}. The question for the maximum of several random variables is a
classical one, treated in the statistics community under the term \emph{extreme
value
statistics}~\cite{HaanFerreira06,Resnick87,Coles2001,scipy_genextreme_doc}.

The task can be solved numerically: We generate a set of uniformly random input
vertices, calculate the Delaunay tessellation and its dual. We then measure the
distance between Voronoi vertex and Delaunay vertices in each Delaunay cell,
and determine the largest such distance in the tessellation. In order to obtain
good statistics, we repeat the process for many tessellations which differ by
their random positions of input vertices, keeping the number of Delaunay input
points the same. Finite-size effects are limited to a minimum by performing the
tessellations in a periodic cube of side-length~$\boxlen$.
\begin{figure}%
  \centering
  \includegraphics{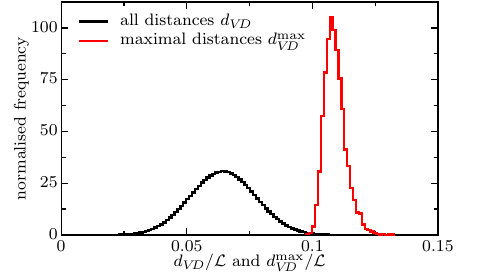}%
  \caption{Black line: Histogram of the distances~$\distDV$ from every Voronoi
  vertex to the corners of its dual Delaunay cell, taken from
  $10^3$~tessellations of uniformly random points in a periodic cube of
  side-length~$\boxlen$. Red line: Histogram of the largest
  distance~$\maxdistDV$ in each of $10^4$~tessellations. In both cases the
  density of input Delaunay vertices is $N_s'=2387/\boxlen^3$. Both histograms
  are \resub{normalized} to have unit integral.}\label{fig:extreme}
\end{figure}%

Figure~\ref{fig:extreme} shows the resulting histograms of both types of
distances. The black line is the histogram of all Delaunay-to-Voronoi
distances~$\distDV$, collected from all distances in many tessellations. Its
shape is close to the probability density function of a normal distribution,
but its lower tail is of course limited by $\distDV>0$, and its upper tail by
the periodic box. The red line in figure~\ref{fig:extreme} is the histogram of
the respective largest such distance found in each tessellation, $\maxdistDV$,
collected from many tessellations. This histogram is close to a
\emph{\resub{generalized} extreme-value distribution}~\cite{scipy_genextreme_doc} which
is \resub{parameterized} by a shape parameter~$c>0$, a location parameter~$\ell$, and a
scaling parameter~$s$, and which has the probability density
function~\cite{scipy_genextreme_ref}
\begin{equation}
  \label{eq:genextreme_pdf}
  \rho(d;c,\ell,s) =
    \frac{1}{s}
    \exp\Bigl(-\Bigl[1-\frac{d{-}\ell}{s/c}\Bigr]^{\frac{1}{c}}\Bigr) \,
    \Bigl[1-\frac{d{-}\ell}{s/c}\Bigr]^{\frac{1}{c}-1}.
\end{equation}
The mean is given in terms of the gamma
function~\cite{scipy_gamma_doc,scipy_genextreme_mean_code},
\begin{equation}
  \label{eq:genextreme_mean}
  \langle d\rangle = \ell + \frac{s}{c}\bigl(1-|\Gamma(c+1)|\bigr).
\end{equation}
\begin{figure}%
  \centering
  \includegraphics{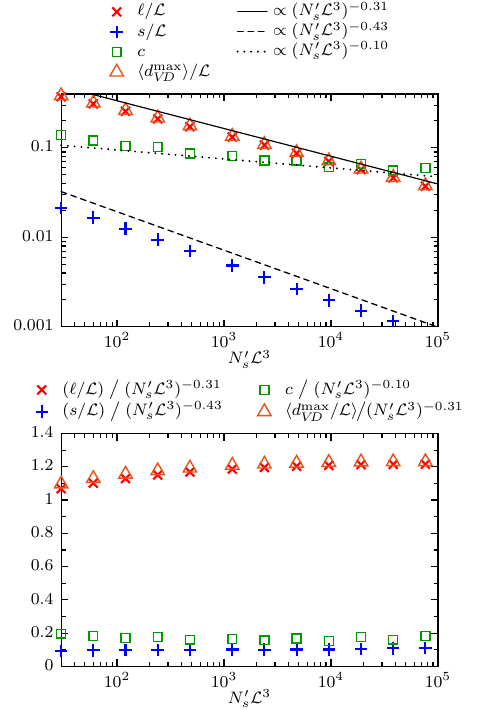}%
  \caption{Dependence on the input density~$N_s'$ of the three
  parameters~$\ell,s,c$ of the \resub{generalized} extreme-value distribution, together
  with its mean value~$\langle\maxdistDV\rangle$. The scaling exponents are
  fitted by the eye.}\label{fig:extreme_scaling}
\end{figure}%
It remains to determine the dependence of parameters $c,\ell,s$ on the
density~$N_s'$ of input points (spherulite centers). We varied $N_s'$ and
fitted the function~\eqref{eq:genextreme_pdf} to the observed
distances~$\maxdistDV$. The best-fit parameters are given in
figure~\ref{fig:extreme_scaling}. In the upper graph we observe that the
parameters follow power laws, and we determine their exponents. In the lower
graph we determine the prefactors. The mean of the maximal distance then scales
as
\begin{equation}
  \label{eq:extreme_dist}
  \langle\maxdistDV\rangle \approx 1.2\boxlen\,(N_s'\boxlen^3)^{-0.31},
\end{equation}
with better matches for large~$N_s'$. With expression~\eqref{eq:extreme_dist}
we obtain the average time~$\tau_f$ when the last little cavity of amorphous
matrix is being converted into spherulite material,
\begin{align}
  \label{eq:taur_orig}
  \tau_f
    &= \min\Bigl\{\frac{1}{\dotR_s} \Bigl(\bigl\langle\maxdistDV\bigr\rangle - R_{s0}\Bigr), \tau_h\Bigr\} \\
  \label{eq:taur}
    &\approx \min\Bigl\{\frac{1}{\dotR_s} \Bigl(1.2\boxlen\,(N_s'\boxlen^3)^{-0.31} - R_{s0}\Bigr), \tau_h\Bigr\}.
\end{align}
As argued at the beginning of the present section, this is close to the moment
we called \emph{release of the final (big) cluster} in the previous sections,
and which implies the end of the depolymerization process, and which is marked
by the last big step in figure~\ref{fig:release}. Inserting the parameters of
figure~\ref{fig:release}, we obtain the values
\begin{equation}
  \iflowquality
\else
    \def\hsep{\rule{5pt}{0pt}}
    \def\vsep{\rule{0pt}{10pt}}
    \begin{array}{r<{\hsep}|>{\hsep}c}
                          & \tau_f/\tau_h \\\hline
      \text{dilute case}  & 1.00 \vsep\\
      \text{diverse case} & 0.86 \\
      \text{dense case}   & 0.11 \rlap{.}
    \end{array}
  \fi
\end{equation}
These values are also marked on the upper time axes of
figure~\ref{fig:release}. In the ``dilute'' case the formula~\eqref{eq:taur}
hits the maximal allowed value, $\tau_h$, which is the correct \resub{behavior} also
observed in figure~\ref{fig:release}. In the ``diverse'' case, the formula
slightly overestimates the release time -- either because the latter is
stochastic in nature and figure~\ref{fig:release} shows just one \resub{realization},
or because the difference between the first closure of the amorphous matrix'
exposed surface and the last filling of its bulk by spherulites is not as small
as we expected. In the ``dense'' case, the formula correctly predicts the
release time of the cluster.

\section{\resub{Example: Predictions for PET waste depolymerization}}
\label{sec:exp}
%
\begin{figure}%
  \centering
  \includegraphics{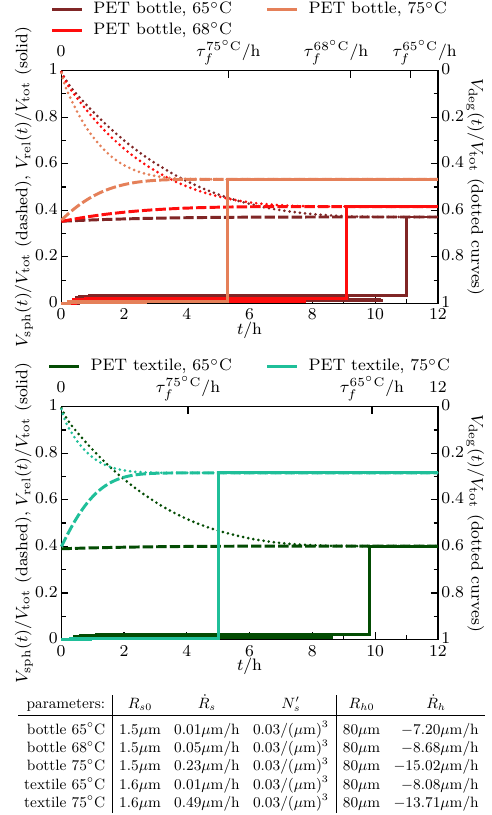}
  \caption{Release of the spherulite clusters (continuous lines), and evolution
  in time of the four volumes $\Vsphrel(t)$, $\Vsphatt(t)$, $\Vam(t)$,
  $\Vdeg(t)$. The plot style is the same as in figure~\ref{fig:release}:
  $\Vdeg(t)$ (dotted curves) refer to the right axis; $\Vsph(t)$ (dashed
  curves) and $\Vsphrel(t)$ (solid lines) refer to the left axis. The colors
  \resub{refer to different} materials. The model parameters used for the graphs, as
  given in the table, are obtained from experimental data from two different
  waste sources at different temperatures, see reference~\textcite{ourAvrami}.}%
  \label{fig:cgal_exp_release}
\end{figure}%
\begin{figure}%
  \centering
  \includegraphics{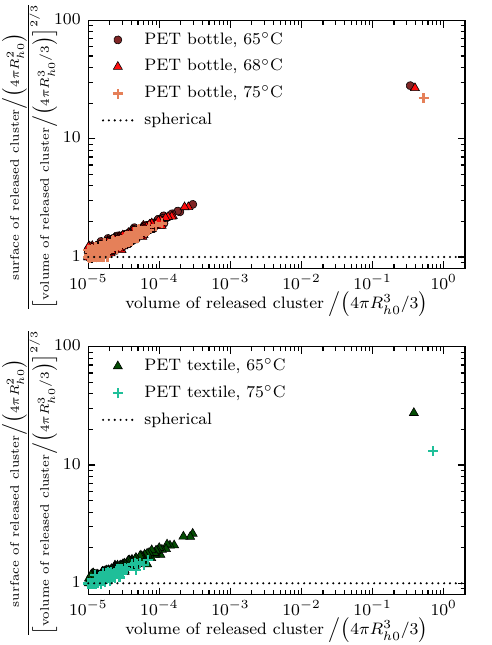}
  \caption{Geometric measures of every spherulite cluster released in the
  evolutions shown in figure~\ref{fig:cgal_exp_release}. Numeric values are
  given in table~\ref{tab:cgal_exp_clusters}.}%
  \label{fig:cgal_exp_clusters}
\end{figure}%
\begin{table*}%
  \def\hsep{\rule{5pt}{0pt}}
  \def\vsep{\rule{0pt}{12pt}}
  \centering
  \iflowquality
\else
    \begin{tabular}{r<{\hsep}|c|c|c|c|c}
     overall quantities: & bottle $65^\circ\celsius$ & bottle $68^\circ\celsius$ & bottle $75^\circ\celsius$ & textile $65^\circ\celsius$ & textile $75^\circ\celsius$ \\\hline
      $N_s$ & 63811 & 63694 & 64038 & 63681 & 64138 \vsep\\
     number of clusters & 2429 & 1539 & 771 & 1474 & 267 \vsep\\
     $\displaystyle\Bigl(\sum\limits_c\text{volume of cluster}~c\Bigr)\Bigm/\bigl(4\pi R_{h0}^3/3\bigr)$ & 0.37 & 0.42 & 0.53 & 0.40 & 0.72 \\
     $\displaystyle\Bigl(\sum\limits_c\text{surface of cluster}~c\Bigr)\Bigm/\bigl(4\pi R_{h0}^2\bigr)$ & 15.2 & 15.4 & 14.8 & 15.4 & 10.7 \\
     $\displaystyle\frac{\sum_c\text{surface of cluster}~c}{4\pi R_{h0}^2} \Bigm/
                        {\Bigl[\sum_c\frac{\text{volume of cluster}~c}{4\pi R_{h0}^3/3}\Bigr]^{2/3}}$ & 29.5 & 27.7 & 22.5 & 28.4 & 13.4 \\
     \hline
     \rule{0pt}{16pt}quantities of the dominant cluster~$c$: & & & & & \\\hline
     number of spherulites in cluster~$c$ & 58642 & 60506 & 62622 & 60790 & 63732 \vsep\\
     $\displaystyle\bigl(\text{volume of cluster}~c\bigr)
                   \bigm/
                   \bigl(4\pi R_{h0}^3/3\bigr)$ & 0.34 & 0.39 & 0.52 & 0.38 & 0.71 \vsep\\
     $\displaystyle\bigl(\text{surface of cluster}~c\bigr)
                   \bigm/
                   \bigl(4\pi R_{h0}^2\bigr)$ & 13.6 & 14.4 & 14.3 & 14.5 & 10.5 \vsep\\
     $\displaystyle\frac{\text{surface of cluster}~c}{4\pi R_{h0}^2}
                   \Bigm/
                   \Bigl[\sum_c\frac{\text{volume of cluster}~c}{4\pi R_{h0}^3/3}\Bigr]^{2/3}$ & 28.1 & 26.8 & 22.1 & 27.6 & 13.2
    \end{tabular}
  \fi
  \caption{Quantitative measures of the spherulite clusters shown in
  figure~\ref{fig:cgal_exp_clusters}.}%
  \label{tab:cgal_exp_clusters}
\end{table*}%

In reference~\textcite{ourAvrami} we \resub{fitted the model parameters to
experimental data}. \resub{There}, two different sources of PET were
treated, from bottle flakes, and from textile waste~\cite{ourPNAS}, and they
were degraded at different temperatures. Fitting the model to the experimental
data yielded the parameters given in the table of
figure~\ref{fig:cgal_exp_release}. \resub{In particular, the initial sizes of
the particle and of the spherulites are taken to be
\begin{equation}
  \begin{aligned}
    R_{h0} &= 80\micrometer, \\
    R_{s0} &= 1.5\micrometer \quad\text{or}\quad 1.6\micrometer.
  \end{aligned}
\end{equation}
}

The solution of the model equations for these parameters are reproduced in the
graphs of figure~\ref{fig:cgal_exp_release}. The style of the graphs is the
same as of those in figure~\ref{fig:release}, showing the four volumes
$\Vdeg(t)$, $\Vam(t)$, $\Vsphatt(t)$, $\Vsphrel(t)$. For a detailed
interpretation of these curves see reference~\textcite{ourAvrami}.

We can now add the \resub{numerical prediction for} the remaining spherulite clusters to the
expected outcome of these experiments. The release curves of spherulite \resub{clusters,
obtained} from detailed numerics, are shown in the graphs of
figure~\ref{fig:cgal_exp_release} as continuous lines. The \resub{formation of spherulite
clusters} has two characteristic times, as discussed in
sections~\ref{sec:percolation} and~\ref{sec:extreme}, namely $\tau_p$~for the
occurrence of a globally connected dominant cluster, and $\tau_f$ of the final
release of the remaining cluster.
\iflowquality
\else
We obtain their values from equations~\eqref{eq:taup} and~\eqref{eq:taur},
\begin{equation}
  \label{eq:exp_tau}
  \def\hsep{\rule{5pt}{0pt}}
  \def\vsep{\rule{0pt}{10pt}}
  \def\pz{\phantom{0}}
  \begin{array}{r<{\hsep}|>{\hsep}cccc}
                                      &    \tau_h   &      \tau_p     &     \tau_f  & \tau_f/\tau_h\\\hline
    \text{bottle $65^\circ\celsius$}  &   11.1\hour &  \pz{-9.3}\hour &   11.1\hour & 1\phantom{.0} \vsep\\
    \text{bottle $68^\circ\celsius$}  & \pz9.2\hour &  \pz{-2.2}\hour & \pz9.2\hour & 1\phantom{.0} \\
    \text{bottle $75^\circ\celsius$}  & \pz5.3\hour &  \pz{-0.5}\hour & \pz5.3\hour & 1\phantom{.0} \\
    \text{textile $65^\circ\celsius$} & \pz9.9\hour &    {-22.2}\hour & \pz9.9\hour & 1\phantom{.0} \\
    \text{textile $75^\circ\celsius$} & \pz5.8\hour &  \pz{-0.4}\hour & \pz4.0\hour & 0.7\rlap{\;.}
  \end{array}
\end{equation}
\fi
The values for the percolation time~$\tau_p$ are all negative, which means that
the initial parameters are already above the percolation threshold. It is
therefore impossible to find here \resub{the ``dilute case'' that we saw in}
figures~\ref{fig:release}, \ref{fig:cluster_measures}, and
\ref{fig:cluster_sizes}. The experimental conditions used on the given materials
always lead to a dominant remaining cluster of spherulites, either to a densely
packed one or to a fluffy one. There are also additional small clusters.

The characteristic times in
\iflowquality
\else
equation~\eqref{eq:exp_tau} are marked also on the
\fi
upper time axes in the graphs of figure~\ref{fig:cgal_exp_release}. They
coincide very well with the release times of the final particle, which are
obtained from the detailed numerical solution of the model, and which is
visible as the big steps in the curves indicating the cluster releases (thick
continuous lines). This coincidence is not a surprise, as
\iflowquality
\else
equation~\eqref{eq:exp_tau}
\fi
gives nearly all release times as the largest possible
ones, $\tau_h$, with one exception, $\tau_f^{75^\circ\celsius}$. We thus learn
that the experimental conditions were such that the final clusters were
released at the end of the degradation process, with the degradable matrix
remaining accessible to enzymes down to the center of the particle. The
resulting clusters of spherulites must have a fluffy structure with deep holes
down to the center.

Figure~\ref{fig:cgal_exp_clusters} corroborates this conclusion. It shows the
surface and volume of all spherulite clusters obtained. In all five cases we
see a clear separation of clusters into one dominant and into very small
clusters. The dominant cluster is always made of around $90\%$ of the
spherulites, and it always has a high or very high surface-to-volume ratio,
which is characteristic for a fluffy, loosely connected structure. The numeric
values of these ratios are given in the bottom row of
table~\ref{tab:cgal_exp_clusters}.

\resub{We remind the reader that the numbers presented in
table~\ref{tab:cgal_exp_clusters} and in the figures result from numerical
modeling. We rely on the model parameters we have fitted to experimental data
in reference~\textcite{ourAvrami}. What does our prediction in
table~\ref{tab:cgal_exp_clusters} imply in terms of particles per liter? In
reference~\textcite{ourPNAS} we used around $5\milligramm/\milliliter$ of
polyester powder, up to size $150\micrometer$, in $1\milliliter$ of buffer
solution. Thus, there were around~$1.7\times10^3$ particles in $1\milliliter$
solution, initially. According to table~\ref{tab:cgal_exp_clusters}, for
example for PET bottles processed at $65^\circ\celsius$, these particles are
reduced to~$34\%$ of their initial volume, and around $4.1\times10^6$
additional small particles are split off from them. These millions of small
particles represent only $3\%$~of the initial volume.}

\section{Conclusion}
\label{sec:conclusion}
\resub{For enzymatic depolymerization, plastic material is thermally
pre-treated and mechanically broken into amorphous particles. The process is
then a competition between depolymerization at the surface and growth of
recalcitrant embedded spherulites. Under typical conditions, the undegraded,
residual material represents a significant fraction of the initial polymer
mass. Depending on the quality of the waste stream and its treatment history,
at least 40\% of the material actually remains in form of spherulitic
clusters.}

\resub{Here, we characterize the morphology of these spherulitic clusters.}
We extended the numerics of our geometric model for enzymatic degradation of
plastic particles from reference~\textcite{ourAvrami}, such that it yields also
the clusters of connected crystalline spherulites. The \resub{numerics make}
extensive use of Delaunay/Voronoi tessellations.

In the degradation dynamics of plastic particles we identify three prototypical
cases which differ in the quantity and in the \resub{properties} of the
resulting spherulite clusters. They are presented in figures~\ref{fig:release},
\ref{fig:cluster_measures} \resub{and \ref{fig:cluster_sizes}}: If the spherulites are
few and grow slowly, they form only small clusters. At intermediate growth
rate, called the ``diverse case'', they form both small clusters and a dominant
one. The latter cluster is a loosely connected, fluffy structure of many
spherulites. It has holes and crevasses nearly down to its center. It is
further \resub{characterized} by a very large surface-to-volume ratio. If the
spherulites' growth rate is high, they all form a single, filled cluster with a
low surface-to-volume ratio close to that of a sphere. We provide quantitative
measures of these clusters, both their overall quantities, and also those of
the largest/dominant cluster.

We find that the formation of spherulite clusters is \resub{characterized} by two
timescales, one for the formation of a dominant cluster, the other for the
formation of a dense core in such a cluster. We provide analytic estimates for
both timescales -- for each of them we focus the model to a simpler, effective
one that allows for an alternative treatment. One of them is similar to a
percolation model, well-studied in the statistical physics of disordered media,
the other one makes use of Delaunay/Voronoi tessellations and extreme-value
statistics.

We apply the model to parameters fitted to and matching the outcome of
experiments on the degradation of PET from two different sources of waste,
treated at different temperatures. The degradation dynamics and the resulting
spherulites are presented in figures~\ref{fig:cgal_exp_release}
and~\ref{fig:cgal_exp_clusters}. We find that under the given experimental
conditions the degradation was always in the ``diverse case'', leading to a
dominant remaining cluster of spherulites. The ``dilute case'' is excluded by
means of the high initial crystallinity. The ``dense case'' would also be
possible, but the growth rates of spherulites turn out to be small enough for
this case not to occur.

\resub{The model predicts the resulting semi-crystalline particles as fluffy
assemblies of small spherulites, which have practically no solid core
and a very high surface-to-volume ratio. Their size is nearly the size
of the initial particle, which results from the grinding pre-treatment; in the
example examined in this paper the initial particle size was
around~$160\micrometer$. In addition to the dominant cluster, the
depolymerization produces very small spherulite clusters. In our case of PET
these small microparticles have sizes between $3\micrometer$
and~$30\micrometer$. Together they make up only a few percent of the initial
volume, see table~\ref{tab:cgal_exp_clusters}. However, the number of these
small clusters is huge: In our example of PET depolymerization, they are three
orders of magnitude more numerous than the initial particles (between $267$ and
$2429$, see table~\ref{tab:cgal_exp_clusters}). The exact numbers depend on the
material, they are influenced by the pre-treatment, and they depend on the
nucleation rate and growth rate of spherulites under depolymerization reaction
conditions.}

\resub{The presented model is intended to draw attention to the microparticle
production during the depolymerization process and to help to find
efficient strategies to deal with this leftover material in the overall design
of the recycling process.}

\bibliography{clusters}

\end{document}